# Quantum Sensing of Spin Transport Properties of an Antiferromagnetic Insulator

Hailong Wang[1], Shu Zhang[2], Nathan J. McLaughlin[3], Benedetta Flebus[4,5], Mengqi Huang[3], Yuxuan Xiao[1], Eric E. Fullerton[1], Yaroslav Tserkovnyak[2], Chunhui Rita Du[1,3]

[1]Center for Memory and Recording Research, University of California, San Diego, La Jolla, California 92093-0401
[2]Department of Physics and Astronomy, University of California, Los Angeles, California 90095
[3]Department of Physics, University of California, San Diego, La Jolla, California 92093
[4]Department of Physics, The University of Texas at Austin, Austin, Texas 78712
[5]Department of Physics, Boston College, Chestnut Hill, Massachusetts 02467

Antiferromagnetic insulators (AFIs) are of significant interest due to their potential to develop next-generation spintronic devices. One major effort in this emerging field is to harness AFIs for long-range spin information communication and storage. Here, we report a non-invasive method to optically access the intrinsic spin transport properties of an archetypical AFI $\alpha$-Fe$_2$O$_3$ via nitrogen-vacancy (NV) quantum spin sensors. By NV relaxometry measurements, we successfully detect the time-dependent fluctuations of the longitudinal spin density of $\alpha$-Fe$_2$O$_3$. The observed frequency dependence of the NV relaxation rate is in agreement with a theoretical model, from which an intrinsic spin diffusion constant of $\alpha$-Fe$_2$O$_3$ is experimentally measured in the absence of external spin biases. Our results highlight the significant opportunity offered by NV centers in diagnosing the underlying spin transport properties in a broad range of high-frequency magnetic materials, which are challenging to access by more conventional measurement techniques.

Advanced materials are integral to both scientific research and modern technology enabling a wide range of emerging applications. Antiferromagnetic insulators (AFIs), a group of scientifically intriguing and technologically relevant materials, promise to bring new functionalities, such as high-density magnetic memory, ultrafast data processing speeds, and robust strategies against malignant perturbations to develop transformative spin-based information technologies.[1–3] Many of these advantages derive from their exchange-enhanced magnon band gaps reaching into a terahertz (THz) regime and the vanishing net magnetization of the Néel order, which enable opportunities for outperforming their ferromagnetic counterparts.[4–8]

Recent efforts have pursued effective control and manipulation of the Néel orientation in a range of AFIs.[9–14] Building on the excellent spin coherence and reduced energy dissipation channels, AFIs also stand out as an attractive solid-state medium to support long-range transmission of spin information in an energy-efficient manner.[15–17] In the present state of the art, spin transport in AFIs is typically investigated via transport measurements in a non-local geometry, in which two patterned heavy metal strips with a large spin-orbit coupling serve as a spin injector and detector via the spin Hall effect and inverse spin Hall effect, respectively. The AFI embedded between the two metallic contacts provides the spin transport channel.[15,16,18,19] The characteristic spin diffusion length is obtained by measuring the spatially-dependent variation of the longitudinal spin accumulation. An essential element in the non-local measurement scheme is an external spin bias established either by electrical spin Hall injection or temperature gradient,[20,21,22] which is usually sensitive to interfacial/surface conditions and is challenging to precisely control in miniaturized spintronic devices.

In this Letter, we report a non-invasive measurement technique that takes advantage of nitrogen-vacancy (NV) centers to reveal the intrinsic spin transport properties of an archetypical AFI $\alpha$-Fe$_2$O$_3$.[10,15,19] By measuring the longitudinal spin noise associated with energy transfer between two magnon states, NV centers access the fluctuation of the longitudinal spin density in the time domain, offering an alternative way to extract the intrinsic spin diffusion constant in the absence of external spin stimuli. We emphasize that the probing frequency of the NV quantum sensing platform is well below the AF magnon band minimum, suggesting that this approach can be extended naturally to many other high-frequency magnetic systems, such as two-dimensional magnets,[23,24] spin liquids,[25] and magnetic Weyl semimetals,[26] whose spin transport behaviors are challenging to access by more conventional magnetic resonance techniques.

We first briefly review the pertinent quantum-mechanical properties of NV centers. An NV center is formed by a nitrogen atom adjacent to a carbon atom vacancy in one of the nearest neighboring sites of a diamond crystal lattice.[27] The negatively charged NV state has an $S = 1$ electron spin and naturally serves as a three-level qubit system. Due to their excellent quantum coherence, single-spin sensitivity, and notable versatility in a broad temperature range,[27,28] NV centers have been demonstrated to be a transformative tool in exploring magnetic and electric properties in a variety of quantum materials with unprecedented field sensitivity and nanoscale spatial resolution.[29–31] In our measurements, we transferred patterned diamond nanobeams containing individually addressable NV centers on the surface of an $\alpha$-Fe$_2$O$_3$ single crystal, as illustrated in Fig. 1(a). The diamond nanobeam has the shape of an equilateral triangular prism with dimensions of 500 nm × 500 nm × 10 μm to ensure nanoscale proximity between NV centers and the sample studied (see supplementary information for details).[32–34] A 200-nm-thick Au stripline was fabricated on the $\alpha$-Fe$_2$O$_3$ crystal to provide microwave control of the NV spin states.[35] An external magnetic field $H$ was applied and aligned to the NV-axis. A photoluminescence (PL) image shows two individual NV centers positioned on top of the $\alpha$-Fe$_2$O$_3$ crystal [Fig. 1(b)], demonstrating the single-spin addressability of the measurement platform.

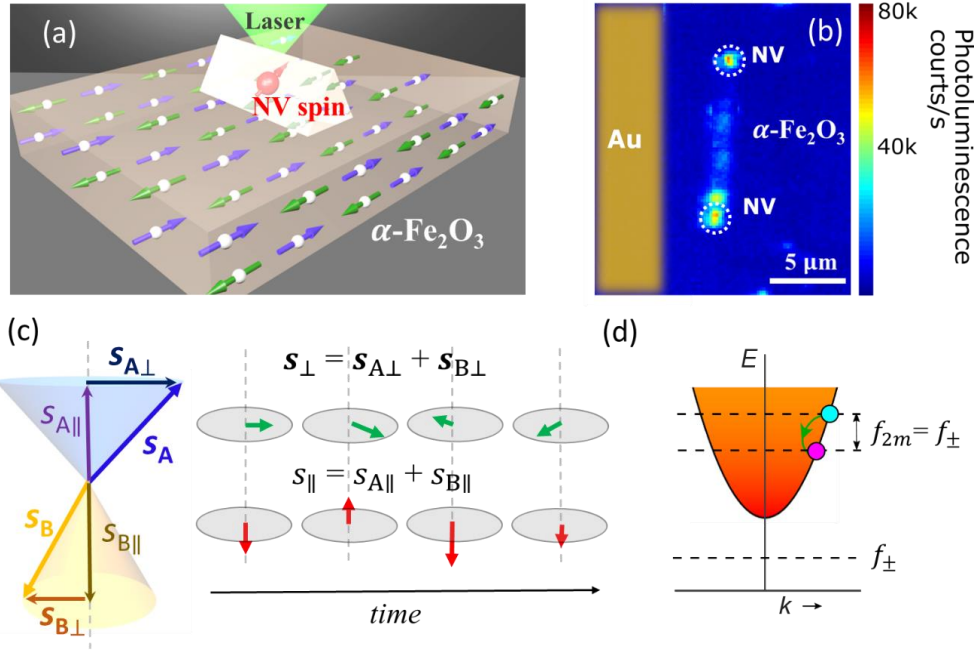

**Figure 1**. (a) Schematic of a single NV spin contained in a patterned diamond nanobeam locally probing the longitudinal spin noise generated by an $\alpha$-Fe$_2$O$_3$ crystal. (b) A photoluminescence image showing a diamond nanobeam containing two individually addressable NV spins positioned on top of an $\alpha$-Fe$_2$O$_3$ crystal. (c) Left: spin configuration of a Néel vector of a bipartite antiferromagnet. The dynamical opening angle of the two sublattices is exaggerated for clarity. $s_A$ and $s_B$ represent the spin density of the two AF sublattices. $s_{A\perp}$ ($s_{B\perp}$) and $s_{A\parallel}$ ($s_{B\parallel}$) are the transverse and longitudinal components, respectively. Right: schematic of the time dependent fluctuations of the longitudinal and transverse spin density $s_\parallel$ and $s_\perp$ of the Néel order. (d) Sketch of the magnon dispersion and the magnon density of $\alpha$-Fe$_2$O$_3$, which falls off as 1/energy (1/$E$), as indicated by the fading colors. In the low field regime, the frequency gap between the two branches of AF magnon modes is negligible compared with the AF magnon band gap. The longitudinal spin fluctuations are associated with two-magnon scattering processes, which are coupled with an NV center through dipolar interactions. $f_\pm$ correspond to the NV electron spin resonance frequencies and $f_{2m}$ represents the frequency of the longitudinal spin fluctuation.

$\alpha$-Fe$_2$O$_3$ exhibits a characteristic Morin phase transition at a critical temperature ($T_M$) of ~ 263 K, where it changes from an easy-plane, canted AFI to a conventional uniaxial AFI with an easy-axis along the [0001]-crystalline orientation.[15,17] We first focus on the uniaxial AF state with a temperature below $T_M$. Figure 1(c) illustrates the time dependent fluctuations of the transverse and longitudinal spin density $s_\perp$ and $s_\parallel$ of the Néel vector of $\alpha$-Fe$_2$O$_3$ at thermal equilibrium. The dynamical opening angle of the two sublattices is exaggerated for clarity. The local fluctuations of $s_\perp(t)$ and $s_\parallel(t)$ will generate the spin noise $\delta \mathbf{B}_\perp(t)$ and $\delta \mathbf{B}_\parallel(t)$, respectively. Such magnetic noise can be understood by invoking the Holstein-Primakoff transformation: $s_\perp \propto \alpha^+(\alpha)$, $s_\parallel \propto s - \alpha^+ \alpha$ with $\alpha^+(\alpha)$ being the magnon creation (annihilation) operator.[36] In an intuitive picture, the transverse spin noise $\delta \mathbf{B}_\perp(t)$ is related to one-magnon scattering processes,[33,37,38] *i.e.* the creation or annihilation of magnons, which vanish at frequencies below the magnon band minimum. The longitudinal spin noise $\delta \mathbf{B}_\parallel(t)$ is related to two-magnon scattering processes, where a magnon with a frequency $f + f_{2m}$ can undergo a transition to another magnon with a frequency $f$, emitting magnetic noise at frequency $f_{2m}$ as illustrated in Fig. 1(d).[37,38] Microscopically, the time dependent fluctuations of the longitudinal spin density $s_\parallel$ reflect spin transport via Brownian motion and naturally connect to the intrinsic spin diffusion constant $D$ through the conventional diffusion equation:[37,39]

$$\partial_t s_\parallel - D \nabla^2 s_\parallel = -\frac{1}{\tau_s} s_\parallel \tag{1}$$

where $\tau_s$ is the coarse-grained spin relaxation time. Figure 1(d) illustrates the magnon band structure of $\alpha$-Fe$_2$O$_3$ and the generation of longitudinal spin noise. In contrast to the transverse counterpart with a minimal frequency determined by magnon band gap, the frequency of the longitudinal spin noise starts from zero and extends to the available magnon band energy,[37] which can be addressed by NV centers in the low GHz regime.

We employed NV relaxometry measurements[27,33,40,41] to detect the longitudinal spin noise generated by the $\alpha$-Fe$_2$O$_3$ crystal. The top panel of Fig. 2(a) shows the optical and microwave measurement sequence. A 3-μs-long green laser pulse is first applied to initialize the NV spin to the m$_s$ = 0 state. Dipolar interactions enable coupling between the NV centers and spin fluctuations in the $\alpha$-Fe$_2$O$_3$ crystal. Longitudinal spin noise at frequencies $f_\pm$ induces NV spin transitions from the m$_s$ = 0 to the m$_s$ = ±1 states, leading to an enhancement in the NV relaxation rates $\Gamma_\pm$. Here, $f_\pm$ and $\Gamma_\pm$ correspond to the NV electron spin resonance (ESR) frequencies and NV relaxation rates of the m$_s$ = 0 ↔ ±1 transitions, respectively. Compared with the m$_s$ = 0 state, the optically excited m$_s$ = ±1 states are more likely to be trapped by a dark pathway through an intersystem crossing and back to the m$_s$ = 0 ground state, exhibiting a reduced PL. After a delay time *t*, we measure the occupation probabilities of the NV spin at the m$_s$ = 0 and the m$_s$ = ±1 states by applying a microwave π pulse on the corresponding ESR frequencies and measuring the spin-dependent PL during the first ~600 ns of the green-laser readout pulse. By measuring the integrated PL intensity as a function of the delay time *t* and fitting the data with a three-level model,[33] NV relaxation rates can be quantitatively obtained (see supplementary information for details).

Invoking perturbation theory, the measured NV relaxation rate $\Gamma_\pm$ induced by the longitudinal spin noise at the NV ESR frequencies $f_\pm$ can be expressed as follows:[42]

$$\Gamma_\pm(f_\pm) = \frac{|\gamma \delta B_\parallel(f_\pm)|^2}{2} \tag{2}$$

where $|\delta B_\parallel(f_\pm)|^2$ is the spectral density of the longitudinal spin noise and $\gamma$ is the gyromagnetic ratio of the magnetic sample. According to the fluctuation-dissipation theorem, $|\delta B_\parallel(f_\pm)|^2$ is related to the imaginary part of the longitudinal dynamic spin susceptibility $\chi''$, which can be further expressed as a function of the intrinsic spin diffusion constant $D$ via Eq. (1). In the high temperature limit appropriate for our NV measurements, $\Gamma_\pm$ can be written as (see supplementary information for details):[37]

$$\Gamma_\pm(f_\pm) = \frac{(\gamma\tilde{\gamma})^2 k_B T}{2\pi f_\pm} \int f(\mathbf{k}, d)\chi''(D)\mathrm{d}\mathbf{k} \quad (3)$$

Where $\tilde{\gamma}$ is the gyromagnetic ratio of the NV spin, $k_B$ is the Boltzmann constant, $T$ is temperature, $\mathbf{k}$ is the wavevector of magnons, and $f(\mathbf{k}, d)$ is the transfer function describing the magnetic fields generated at the NV site.[37,42] By setting the frequency at an arbitrary value in Eq. (3), we calculate the longitudinal spin noise spectrum as shown in Fig. 2(b). Because the longitudinal spin noise characterizes the energy transfer between two magnons, qualitatively, a smaller (larger) energy difference, corresponds to a higher (lower) two-magnon scattering rate.

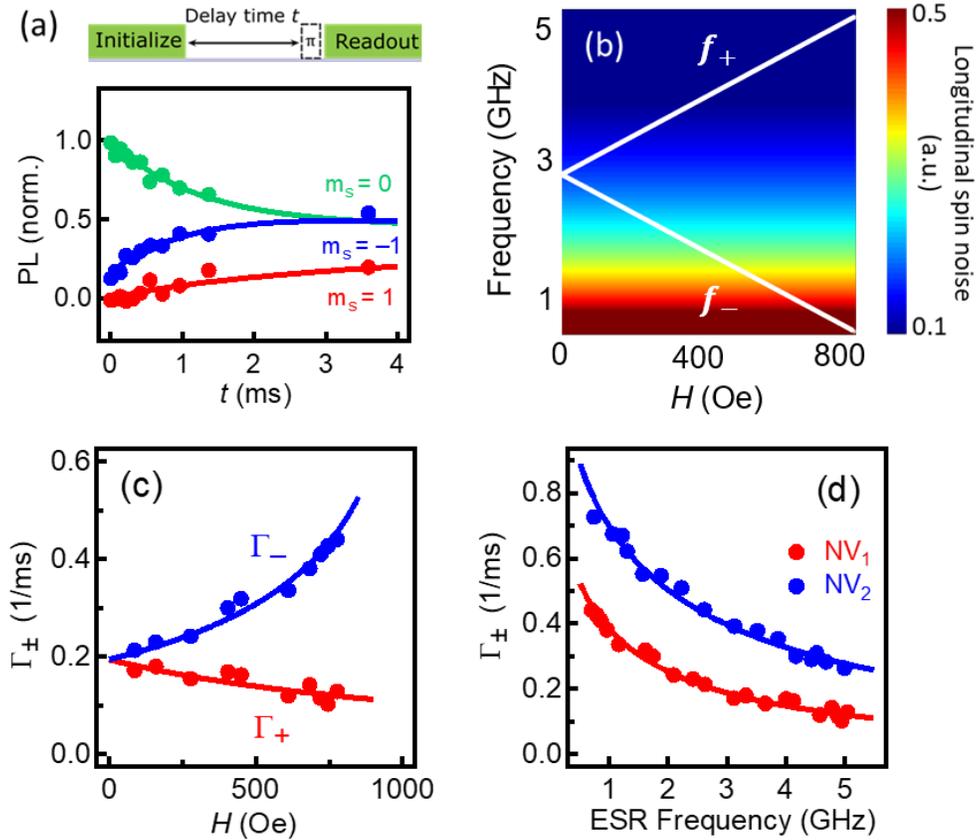

**Figure 2.** (a) Top panel: optical and microwave sequence for NV relaxometry measurements. Bottom panel: A set of NV relaxation data measured with $H$ = 745 Oe. The curves are fitted to a three-level model to extract the NV relaxation rate. (b) Calculated longitudinal spin noise spectrum. The white curves correspond to the characteristic NV ESR frequencies $f_\pm = 2.87 \pm \tilde{\gamma}H/2\pi$. (c) Obtained spin relaxation rates $\Gamma_\pm$ of $NV_1$ as a function of $H$. The red and blue points correspond to $\Gamma_+$ and $\Gamma_-$, respectively. The curves are fitting based on Eq. (3). (d) The spin relaxation rates $\Gamma_\pm$ of $NV_1$ (red dots) and $NV_2$ (blue dots) measured as a function of the NV ESR frequency, which are in agreement with the theoretical model (red and blue curves). The $NV_1$ and $NV_2$ were set at two different distances to the surface of the $\alpha$-$Fe_2O_3$ crystal.

In our experiments, we employed two NV centers ($NV_1$ and $NV_2$) with different NV-to-sample distances ($d_1 = 250 \pm 6$ nm and $d_2 = 185 \pm 5$ nm, see supplementary information for details) to perform the NV relaxometry measurements. The bottom panel of Fig. 2(a) shows a set of spin relaxation data measured on $NV_1$ when $H = 745$ Oe and $T = 200$ K. The measured PL intensity corresponding to the $m_s = 0$ ($\pm 1$) state decreases (increases) as a function of the delay time $t$, indicating a relaxation of the NV spin to a mixture of the $m_s = 0$ and the $m_s = \pm 1$ states. Figure 2(c) shows the extracted NV spin relaxation rate $\Gamma_\pm$ as a function of $H$. The measured magnetic field dependence of $\Gamma_\pm$ is also in agreement with the calculated magnetic noise spectrum as shown in Fig. 2(b).

To quantitatively determine the intrinsic spin diffusion constant of the $\alpha$-$Fe_2O_3$ crystal, we fit the NV ESR frequency dependence of the measured relaxation rates $\Gamma_\pm$ to Eq. (3). The agreement between the experimental results and the theoretical model as shown in Fig. 2(d) provides a strong evidence that we are indeed probing the spin noise generated by the longitudinal spin fluctuation. The intrinsic spin diffusion constant $D$ of $\alpha$-$Fe_2O_3$ is obtained to be $(8.9 \pm 0.5) \times 10^{-4}$ m$^2$/s at 200 K, from which the spin conductivity is calculated to be $(7.1 \pm 0.4) \times 10^6$ S/m (see supplementary information for details). We highlight that the measurements by two NV centers ($NV_1$ and $NV_2$) set at different distances to the sample surface give the same spin diffusion constant of the $\alpha$-$Fe_2O_3$ crystal, which is independent of the details of the NV quantum spin sensors. Taking the measured spin diffusion constant and a magnon velocity $v \sim 30$ km/s,[43] we get momentum scattering time of ~3 ps and a magnon mean free path of ~90 nm for $\alpha$-$Fe_2O_3$. Using spin relaxation time $\tau_s = 10$ ns which is suggested by a rough estimation in Ref. 43,[44] the spin diffusion length $\lambda = \sqrt{D\tau_s}$ of $\alpha$-$Fe_2O_3$ is estimated to be 3 μm at 200 K, in agreement with values reported by the non-local spin transport measurements.[15,17] Note that the magnon mean free path is orders of magnitude smaller than the estimated spin diffusion length, confirming the diffusive nature of AF magnons in $\alpha$-$Fe_2O_3$.

Very recently, non-local spin transport behaviors have been studied in $\alpha$-$Fe_2O_3$ crystals (films) in both the easy-plane and uniaxial AFI phases.[15,19] Next, we investigate the temperature dependence of the intrinsic spin diffusion constant of $\alpha$-$Fe_2O_3$. In particular, we are interested in revealing the variation of spin diffusion constant across the Morin phase transition. Figure 3(a) shows a temperature dependence of the magnetization of the $\alpha$-$Fe_2O_3$ crystal. The external magnetic field is set to be 1000 Oe, perpendicular to the [0001]-crystalline axis. Above the Morin transition temperature ($T_M \sim 263$ K), a weak canted magnetic moment exists in the (0001) magnetic easy-plane, which is confirmed by a field-dependent magnetization curve as shown in the inset. When $T < T_M$, there is a complete absence of a ferromagnetic moment. Figure 3(b) shows the temperature dependence of the spin diffusion constant $D$ measured between 200 K to 300 K, which exhibits a gradual decrease from $(8.9 \pm 0.5) \times 10^{-4}$ to $(6.6 \pm 0.4) \times 10^{-4}$ m$^2$/s. Notably, $D$ smoothly varies across the Morin transition temperature.

Theoretically, the intrinsic spin diffusion constant is determined by the magnon velocity $v \sim \frac{J_s a}{\hbar}$ ($J_s$ is the exchange constant, $a$ is the lattice constant, and $\hbar$ is the reduced Planck constant)[45] and the momentum relaxation time $\tau$ as follows: $D = \frac{v^2 \tau}{3}$.[44] Momentum relaxation time $\tau$ in magnetic crystals (films) is determined by disorder, phonons and Umklapp scattering.[21] As the temperature increases, the inelastic scattering rate starts to increase, leading to a reduced $\tau$ and a spin diffusion constant as shown in Fig. 3(b). Across the Morin transition temperature, the smooth variation of $D$ demonstrates that AF spin transport in equilibrium state is mainly driven by thermal magnons governed by the exchange interaction. The variation of magnetic anisotropies and Dzyaloshinskii-Moriya interaction can be treated as small corrections on the thermal energy scale.[46] We highlight that the NV relaxometry method could further be employed to reveal the intrinsic magnetic phase transition as shown in Fig. 3(c). In general, the measured NV relaxation rate $\Gamma_-$ increases with the temperature indicating that the magnetic noise scales with the magnon density.[33] At $T \sim 263$ K, we observed a sudden jump of $\Gamma_-$, which is attributable to the rotation of the Néel order parameter during the Morin transition [see supplementary information for details].

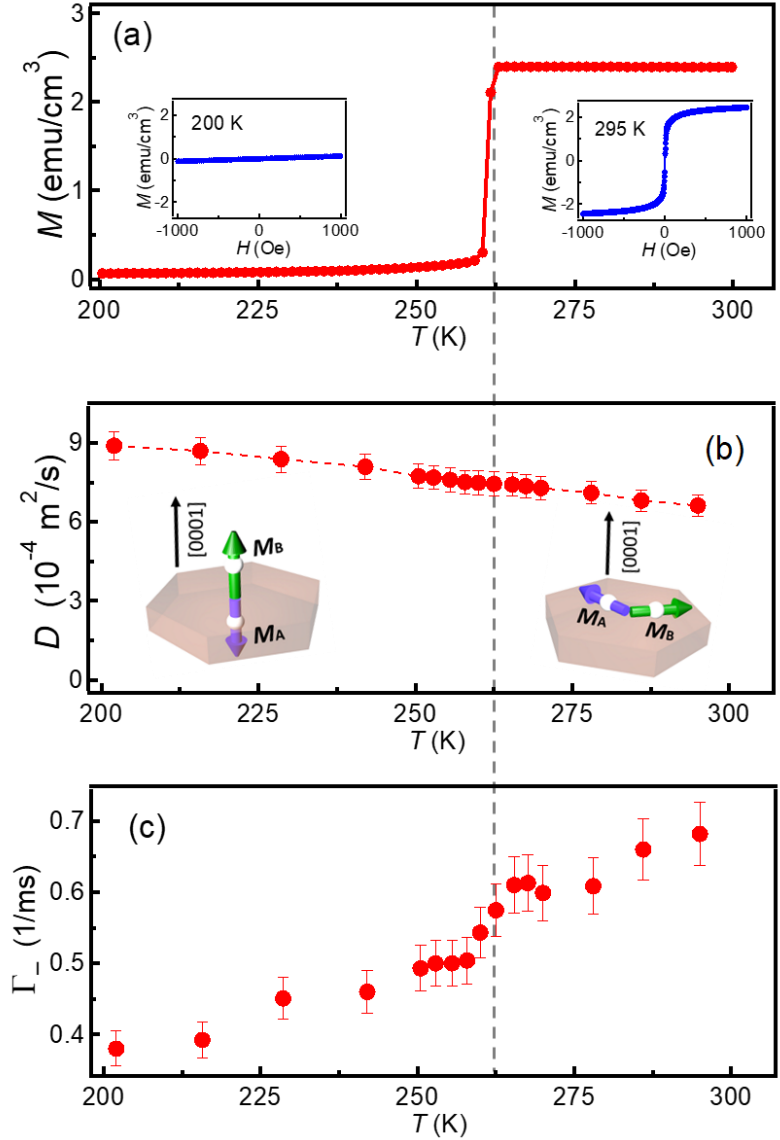

**Figure 3**. (a) Temperature dependence of the magnetization of the $\alpha$-Fe$_2$O$_3$ crystal. Insets: field-dependent magnetization of the $\alpha$-Fe$_2$O$_3$ crystal when $T = 200$ K and 295 K. The dash line represents the Morin transition temperature of $\alpha$-Fe$_2$O$_3$. (b) The intrinsic spin diffusion constant $D$ measured as a function of temperature between 200 K and 300 K. Insets: schematic views of the magnetic order of $\alpha$-Fe$_2$O$_3$ below and above the Morin transition temperature. $M_A$ and $M_B$ illustrate the magnetic moment of the two AF spin sublattices. (c) The NV relaxation rate $\Gamma_-$ measured as a function of temperature. The external magnetic field $H$ is set to be 683 Oe, corresponding to the NV ESR frequency $f_-$ of 0.96 GHz.

In summary, we have demonstrated NV centers as a local probe of intrinsic spin transport properties of a proximal $\alpha$-$Fe_2O_3$ crystal. In contrast to the existing measurement platforms that involve integrating materials into the device structure, NV centers access the spin diffusion constant non-invasively by probing the local (equilibrium) longitudinal spin fluctuations in the time domain, which obviates the requirement of an external spin bias. We highlight that this merit is of particular relevance when it comes to investigating spin transport in high-frequency magnetic materials, whose coherent spin resonances are challenging to excite. By employing scanning NV microscopy or diamond nanopatterns with shallowly implanted NV centers,[30,31] we expect that the spatial resolution of the presented NV quantum sensing platform could ultimately reach the nanometer scale, offering new opportunities to reveal emergent spin transport in a broad class of magnetic systems.

**Acknowledgement:** Authors would like to thank Francesco Casola for fabricating the diamond nanobeams. N. J. M. and C. R. D. were supported by Air Force Office of Scientific Research under award FA9550-20-1-0319. H. W., Y. X., and E. E. F. were supported by the Quantum Materials for Energy Efficient Neuromorphic Computing, an Energy Frontier Research Center funded by the U.S. Department of Energy (DOE), Office of Science, Basic Energy Sciences (BES) under Award DE-SC0019273. S. Z. and Y. T. were supported by NSF under Grant No. DMR-1742928.